%% file: samples/main.tex
\PassOptionsToPackage{table,x11names,dvipsnames}{xcolor}
\documentclass[sigconf]{acmart}
\AtBeginDocument{%
  }

\setcopyright{acmlicensed}
\copyrightyear{2018}
\acmYear{2018}
\acmDOI{XXXXXXX.XXXXXXX}
\acmConference[Conference acronym 'XX]{Make sure to enter the correct
  conference title from your rights confirmation email}{June 03--05,
  2018}{Woodstock, NY}
\acmISBN{978-1-4503-XXXX-X/2018/06}

\usepackage{times}
\usepackage{latexsym}
\usepackage{amsmath}
\usepackage[most]{tcolorbox}
\usepackage{xspace}

\usepackage[utf8]{inputenc}

\usepackage[svgnames]{xcolor}
  \definecolor{diffstart}{named}{Grey}
  \definecolor{diffincl}{named}{Green}
  \definecolor{diffrem}{named}{OrangeRed}

\usepackage{listings}
  \lstdefinelanguage{diff}{
    basicstyle=\small,
    morecomment=[f][\color{diffstart}]{@@},
    morecomment=[f][\color{diffincl}]{+\ },
    morecomment=[f][\color{diffrem}]{-\ },
  }

\usepackage{microtype}

\usepackage{graphicx}
\usepackage{tikz}
\usepackage{booktabs}

\newcommand{\ray}[1]{
}

\newcommand{\ToolName}{\textsc{Refine}\xspace}
\def \tool{\textsc{Refine}\xspace}

\def \tool{\textsc{Refine}\xspace}

\newcommand\figref[1]{Fig.~\ref{#1}}

\newcommand\tabref[1]{Table~\ref{#1}}

\newcommand\secref[1]{\S\ref{#1}}

\newcommand{\fakeparagraph}[1]{\noindent\textbf{#1.}}

\newcommand*\circled[1]{\tikz[baseline=(char.base)]{
            \node[shape=circle,draw,inner sep=0.75pt] (char) {#1};}}
\newcommand*\blackcircled[1]{\tikz[baseline=(char.base)]{
    \node[shape=circle, fill=black, text=white, inner sep=0.75pt, draw=black] (char) {#1};}}

\definecolor{lightmauve}{rgb}{0.86, 0.82, 1.0}

\usepackage{makecell} % Make sure you include this in your preamble if using \makecell
\usepackage{enumitem} %for no space in itemize

\usepackage{circledtext}
\usepackage{newtxtext}
\usepackage{tikz}
\usepackage{libertine}
\usepackage{nicematrix}
\usepackage{multirow} 
\usepackage{cleveref}
\usepackage{url}

\begin{document}

%\title{\ToolName{}: A Semantic Aware Program Repair Agent}
%\title{From Draft to Patch: \\ Hierarchical Agents for Corrective Program Repair}
\title{\tool: Enhancing Program Repair Agents through Context-Aware Patch Refinement}


\author{Anvith Pabba}
\affiliation{%
  \institution{Columbia University}
  % \city{New York}
  \country{}
}
\email{ap4450@columbia.edu}

\author{Simin Chen}
\affiliation{%
  \institution{Columbia University}
  % \city{New York}
  \country{}
}
\email{sc5687@columbia.edu}

\author{Alex Mathai}
\affiliation{%
  \institution{Columbia University}
  % \city{New York}
  \country{}
}
\email{am6215@columbia.edu}

\author{Anindya Chakraborty}
\affiliation{
    \country{}
}
\email{anindyaju99@gmail.com}

\author{Baishakhi Ray}
\affiliation{%
  \institution{Columbia University}
  % \city{New York}
  \country{}
}
\email{rayb@cs.columbia.edu}

\renewcommand{\shortauthors}{Trovato et al.}
\newcommand{\TODO}[1]{\textbf{\color{red}TODO:{ #1} }}
\newcommand{\CM}[1]{\textbf{\color{red}CM:{ #1} }}
\newcommand{\rayb}[1]{\textbf{\color{red}Rayb:{ #1} }}
\newcommand{\ap}[1]{\textbf{\color{cyan}Anvith:{ #1} }}

\begin{abstract}

Large Language Models (LLMs) have recently shown strong potential in automatic program repair (APR), especially in repository-level settings where the goal is to generate patches based on natural language issue descriptions, large codebases, and regression tests. However, despite their promise, current LLM-based APR techniques often struggle to produce correct fixes due to limited understanding of code context and over-reliance on incomplete test suites. As a result, they frequently generate \textsc{Draft Patches}—partially correct patches that either incompletely address the bug or overfit to the test cases. In this work, we propose a novel patch refinement framework, \ToolName{}, that systematically transforms \textsc{Draft Patches} into correct ones. \ToolName{} addresses three key challenges: disambiguating vague issue and code context, diversifying patch candidates through test-time scaling, and aggregating partial fixes via an LLM-powered code review process. We implement \ToolName{} as a general refinement module that can be integrated into both open-agent-based and workflow-based APR systems. Our evaluation on the SWE-Bench Lite benchmark shows that \ToolName{} achieves state-of-the-art results among workflow-based approaches and approaches the best-known performance across all APR categories. Specifically, \ToolName{} boosts AutoCodeRover’s performance by 14.67\%, achieving a score of 51.67\% and surpassing all prior baselines. On SWE-Bench Verified, \ToolName{} improves the resolution rate by 12.2\%, and when integrated across multiple APR systems, it yields an average improvement of 14\%—demonstrating its broad effectiveness and generalizability. These results highlight the effectiveness of refinement as a missing component in current APR pipelines and the potential of agentic collaboration in closing the gap between near-correct and correct patches. We also open source our code here: \href{https://anonymous.4open.science/r/SemAgent-7B2F/README.md}{\textit{Link to Anonymous GitHub Repo}}.
\end{abstract}

% <concept_id>10011007.10011074.10011092.10011782</concept_id>
% <concept_desc>Software and its engineering~Automatic programming</concept_desc>

% \ccsdesc[500]{Software and its engineering~Automatic programming}



\received{20 February 2007}
\received[revised]{12 March 2009}
\received[accepted]{5 June 2009}

% \TODO{GPT 2.5  -> Gemini 2.5}
% \TODO{not use Effective}

\maketitle

\input{samples/1_intro_new}
\input{samples/2_related}

\input{samples/3_motivation}

\input{samples/4_methodology}

\input{samples/5_experiments}

\input{samples/6_results}

\input{samples/7_case_study}

\section{Threats to Validity}

Our reported results may have the following threats to validity.

\paragraph{External Validity.} Our evaluation focuses on the SWE-Bench Lite benchmark, which—while representative—may not fully capture the diversity of bugs and codebases encountered in the wild. To minimize this threat we also report result on SWE-Bench Verified  benchmark.  In particular, the GitHub issues in SWE-Bench are often well-structured and accompanied by relevant test cases, whereas real-world bug reports may be noisier, less detailed, or lack reproducible test environments. The effectiveness of \ToolName{} may vary when applied to such settings.

\paragraph{Construct Validity.} Our formulation of \textsc{Draft Patches}—including \textsc{Incomplete} and \textsc{Overfitted} patches—is grounded in our empirical observations and aligns with known limitations of LLM-based APR systems. However, other categories of failure modes may exist that we do not explicitly model. Additionally, our reliance on regression test outcomes and LLM-based voting for validation introduces potential biases in defining what constitutes a “correct” patch.

\paragraph{Internal Validity.} \ToolName{} depends on several heuristic decisions (e.g., context extraction strategies, delta patch sampling procedures, aggregation rules) that could influence outcomes. While our design aims to generalize across different agents and workflows, we do not perform ablation studies on every component, which could limit interpretability of their individual contributions.

\paragraph{Tooling and Model Dependency.} Our implementation uses specific LLMs (Claude 3.7 and Gemini 2.5 Pro) as underlying agents. Different models may produce qualitatively different behavior, especially in context interpretation and code generation. Our results may therefore not directly generalize to other LLMs or future model versions with different capabilities.

\paragraph{Scalability and Performance.} Although \ToolName{} is designed to work on repository-level repair tasks, its performance may degrade with very large codebases due to context window limits and increased computational cost of test-time scaling. Future work is needed to assess scalability across industrial-scale systems.

Additionally, \ToolName{} shares common risks associated with any APR system deployed in real-world scenarios. These include the injection of vulnerabilities i.e code that introduces new security flaws in the codebase, failure to account for edge cases, and the possibility of unintended consequences in production environments. While our approach tries to minimize errors such as unaccounted edge cases, it remains crucial that the output is thoroughly reviewed before integration without blind overtrust in Autonomy.

\section{Conclusion}

We present \ToolName{}, a context-aware patch refinement framework that significantly enhances automated program repair. \ToolName{} improves AutoCodeRover’s performance by 14.67\% on SWEBench Lite and raises the resolution rate on SWEBench Verified by 12.2\%. When applied across multiple APR systems, \ToolName{} yields consistent gains—improving resolution rates by an average of 14\%, demonstrating its broad effectiveness and generalizability in refining patches.

\bibliographystyle{ACM-Reference-Format}
\bibliography{samples/sample-base}

\end{document}

%% file: samples/1_intro_new.tex
\section{Introduction}

Large Language Models (LLMs) have demonstrated impressive capabilities across a wide range of software engineering tasks. Among these, automatic program repair (APR) stands out as particularly impactful, offering the potential to significantly reduce manual debugging effort, accelerate software maintenance, and enhance overall code quality. Recent advances in LLM-based APR have moved beyond isolated function-level fixes to focus on repository-level repair—a setting that more accurately reflects real-world scenarios, such as resolving user-reported bugs in GitHub repositories. Formally, the repository-level APR task is defined as follows: Given a user-submitted issue description (typically in natural language), the complete source code of the repository, and  a suite of public test cases (usually regression tests), the goal is to automatically synthesize a correct patch by reasoning over the large and complex codebase. The correctness of the generated patch is then validated using the provided regression test suite~\cite{jimenez2024swebenchlanguagemodelsresolve}.

There has been a recent surge in such APR agents, driven by the agentic paradigm in which Large Language Models (LLMs) are treated as autonomous decision-makers augmented with tool use. Under this framework, a growing body of work explores both \textit{open-agent-based}~\cite{wang2024openhands, yang2024sweagentagentcomputerinterfacesenable} and \textit{workflow-based}~\cite{zhang2024autocoderover, ruan2024specrovercodeintentextraction, zhang2024autocoderoverautonomousprogramimprovement} approaches. These methods endow LLMs with tool-augmented capabilities such as searching the codebase~\cite{wang2024openhands, yang2024sweagentagentcomputerinterfacesenable}, editing source files~\cite{zhang2024autocoderover}, executing tests~\cite{xia2024agentlessdemystifyingllmbasedsoftware}, and more. The primary distinction lies in how these paradigms manage the workflow. \textit{Open-agent-based approaches} delegate tool selection and execution to the LLM, allowing it to dynamically decide which tools to invoke and when, based on intermediate feedback. In contrast, \textit{workflow-based approaches} follow a fixed, manually specified sequence of tool invocations, with no adaptive decision-making by the LLM during the repair process.

\paragraph{Challenges.} While recent APR systems show considerable promise, they still struggle to generate high-quality patches for complex, real-world codebases. 
We identify two primary reasons for this limitation. First, current systems often exhibit a \textbf{limited understanding of the broader code context}, resulting in patches that address only part of the underlying issue. Second, they tend to \textbf{rely too heavily on test suites} as the sole signal of correctness. Consequently, patches that pass all tests may still be incorrect or only partially correct, as the test cases often fail to capture the full semantic intent of the issue.

Our initial study shows that these limitations frequently give rise to what we term \textsc{Near-Correct Patches}—patches that are close to being correct but ultimately fall short. We observe two recurring patterns: (1)~\textsc{Incomplete Patches}, which address only a subset of the necessary changes due to shallow reasoning over the code; and (2)~\textsc{Overfitted Patches}, which pass the available tests but fail to generalize, typically because they overfit to weak or underspecified test cases provided in the issue description (see~\Cref{sec:motivation} for examples). 
These partially correct patches are widespread across both open-agent-based and workflow-based APR systems. We collectively refer to these failure modes as \textsc{Draft Patches}—patches that fall short of correctness but contain promising partial signals. Despite their prevalence, current APR systems offer no systematic mechanisms for refining such patches. This gap motivates our work: \textit{to develop principled strategies that can refine \textsc{Draft Patches} into correct and robust fixes.}

Although prior work~\cite{yu2019alleviating,fei2025patch, tianpatch, smith2015cure} has also observed patch overfitting in function-level program repair and proposed training specialized models to iteratively refine incorrect patches into correct ones~\cite{ye2024iter}, these efforts are typically tool-specific. In other words, a refinement method designed for one APR tool cannot be applied to others, which prevents leveraging optimization strategies across different approaches. This limitation becomes even more restrictive in repository-level program repair, where APR agents adopt diverse search strategies, repair paradigms, and patch generation mechanisms. To overcome this, we seek to extend the concept of patch refinement in a generalizable way using LLM agents, enabling a single framework to refine patches across heterogeneous APR tools in a black-box manner.

There are three main challenges in designing a generic patch refinement module.

\begin{enumerate}[leftmargin=*]
  
    \item \textit{Lack of Precise Context.} Refining a \textsc{Draft Patch} requires providing the LLM with rich contextual information. However, natural language issue descriptions are often vague or ambiguous, leading to misinterpretation or omission of key details. Additionally, the required code context for generating a complete patch is often unknown, reducing refinement accuracy. %our solution: context

    \item \textit{Limited Exploration of Delta Patch Space.} Existing methods frequently fail to explore the diverse space of possible delta patches, limiting the chances of discovering more effective or semantically correct refinements. %our solution: test time scale

    \item \textit{Ineffective Selection of Refined Patches.} Even when multiple candidate refinements are produced, current approaches lack principled mechanisms for evaluating and selecting the best ones. This often results in suboptimal patches being chosen, undermining the impact of the refinement process. %our solution: review
\end{enumerate}

To address the aforementioned challenges, we propose the following approach: (1) \textit{Getting the Right Context}: We introduce an agent to disambiguate and enrich both the issue and code context, providing the LLM with clearer and more relevant information necessary for effective patch refinement.
(2) \textit{Diverse Delta Patch Generation}: We apply test-time scaling to generate multiple \textit{Delta Patch} candidates, enabling broader exploration of the solution space and increasing the likelihood of discovering the correct patch.
(3) \textit{Aggregated Patch Synthesis}: We simulate the code review process by introducing a code review agent. This agent aggregates the partially correct fixes from each \textit{Delta Patch} candidate and combines them with the original \textsc{Near-Correct Patch}, ultimately producing a correct and comprehensive patch.

To this end, we implement a novel framework, \ToolName{} that refines \textsc{Draft Patches} generated by APR systems. We first use \tool to refine the patches generated by a variety of APR approaches such as ExpeRepair\cite{li2025experepair}, BlackBoxAI\cite{blackboxai2024}, Agentless\cite{xia2024agentlessdemystifyingllmbasedsoftware}, CodeV\cite{li2024codev}, and AutoCodeRover\cite{zhang2024autocoderover} on a subset of SWEBench lite where our results show that \ToolName{} consistently increases the resolution rate across all approaches on the benchmark.
Next, we perform a full SWEBench Lite \& Verified run with \ToolName{} using AutoCodeRover as the seed, and compare it against a set of baseline APR tools, our results show that we improve the resolution rate of AutoCodeRover on SWEBench Lite by 14.67\% and by 12.2\% on SWEBench Verified, with \ToolName{} out-performing all other baselines on SWEBench Lite.
Finally, we conduct an ablation study to investigate the contribution of each component of \tool{} to its bug-fixing performance. The results demonstrate that every component of \tool{} positively contributes to correcting more \textsc{Draft Patches}. 
% In addition, we observe that neither the choice of backend LLM nor the hyperparameters significantly impact \tool{}'s overall effectiveness.

We summarize our contribution as follows:
\begin{enumerate}[leftmargin=*,noitemsep,topsep=0pt]

\item \textbf{Problem Novelty.} We introduce and investigate a new problem: patch refinement, which aims to iteratively transform \textsc{Draft Patches} into correct ones. Our proposed patch refinement module is designed to be general and can be seamlessly integrated to enhance both agent-based and workflow-based APR techniques.
%\TODO{Confirm whether empirical results support this general claim.}

\item \textbf{Technique Novelty.} We present \ToolName{}, a concrete implementation of patch refinement. Specifically, \ToolName{} consists of three key components: (1) two  agents that get the \texttt{Issue Context} and \texttt{Code Context}, (2) a test-time scaling module that generates diverse plausible \texttt{patch deltas} to expand the search space, and 
(3) a code review agent that aggregates multiple \texttt{patch deltas} from the test-time scaling module to produce correct patches.

\item \textbf{Evaluation.} \ToolName{} demonstrates state-of-the-art performance, when seeded with initial patches from AutoCodeRover - achieving a resolution rate of 51.67\% on SWE-Bench Lite and 63.8\% on SWE-Bench Verified. These results represent an absolute performance increase of 14.67 and 12.2 percentage points, respectively. Furthermore, our evaluation shows that \ToolName{} is effective at refining patches from other leading Automated Program Repair (APR) systems improving their average resolve rate on SWE-Bench Lite by 14\% showing its generalizability and effectiveness in refining patches.

\end{enumerate}

%% file: samples/2_related.tex
\section{Background \& Related Work}
\label{sec:related_work}

\subsection{LLM for Software Engineering}
With the rise of large language models (LLMs), there has been a surge of research on leveraging LLMs for a wide range of downstream software engineering tasks, such as code generation \cite{chen2021evaluatinglargelanguagemodels, chen2024ppm, chen2025dynamic, jain2024livecodebench, ding2024semcoder},  malware detection~\cite{chen2020denas, al2024exploring, jelodar2025large}, test creation \cite{kang2023largelanguagemodelsfewshot, chen2024chatunitest, ryan2024code}, automated code review \cite{li2022automatingcodereviewactivities}, and program repair \cite{gao2022programrepair,li2024hybrid}.

Among these applications, code generation has attracted particular attention from both academia and industry, as evidenced by the emergence of AI-powered development tools like GitHub Copilot, Amazon CodeWhisperer, and Claude Code. Over the past decade, the landscape of code-focused LLMs has rapidly expanded, starting from early models like PaLM to a diverse ecosystem of models spanning a range of sizes and capabilities. Today, this includes powerful closed-source models such as ChatGPT, O3, Claude, and Gemini, as well as open-source alternatives like StarCoder, CodeLlama, Code Gemma, and DeepSeek-Coder, among others.

Beyond code generation, LLMs have demonstrated significant potential across a spectrum of other software engineering tasks. For example, LLMs have been employed for automated test case generation~\cite{zhang2024llm, nan2025test, wang2024hits, xue2024llm4fin, yang2024evaluation}, enabling more thorough and efficient testing processes by suggesting comprehensive and context-aware test scenarios. 
In the domain of code review, LLMs can automatically identify bugs, suggest improvements, and assist in enforcing coding standards, thereby streamlining the review process and improving code quality. 
Other applications include code summarization, documentation generation~\cite{sun2024source, ahmed2022few, crupi2025effectiveness}, code translation~\cite{pan2024lost, chen2024decix, jana2024cotran, luo2024bridging, yan2023codetransocean}, and refactoring, all of which leverage LLMs’ ability to understand and manipulate natural and programming languages. Collectively, these advancements are transforming the software development lifecycle, making it more automated, efficient, and accessible.

\begin{figure*}[t]
    \centering
    \includegraphics[width=0.66\textwidth]{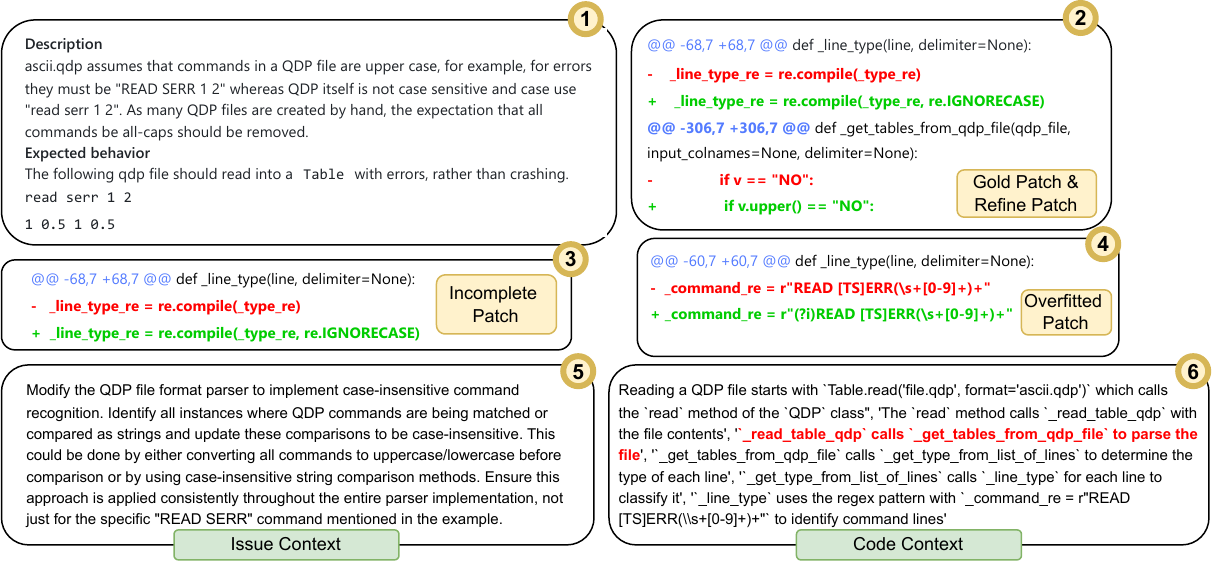}
    \caption[]{Motivating Example - \blackcircled{1} is the issue description, \blackcircled{2} is the developer patch \& \tool patch, \blackcircled{3} and \blackcircled{4} are incomplete/overfitted patches, \blackcircled{5} is the \ToolName{} generated issue semantics, \blackcircled{6} is the \ToolName{} generated code semantics.
    %~\ray{I think 3 is incomplete, not inconsistent----changing the text}
    }
    \label{fig:motivating}
\end{figure*}

\subsection{Automatic Program Repair}

The evolution of APR techniques~\cite{10.1145/3318162} can be traced through several key paradigms: search-based, semantics-based, learning-based, and, most recently, LLM-based approaches.

\fakeparagraph{Traditional APR Techniques} Early research in APR predominantly explored search-based and semantics-based techniques. Search-based methods \cite{5070536} start from a faulty program and apply a predefined set of code mutations to generate candidate patches. These candidates are then validated against a suite of tests, with successful patches being those that pass all relevant test cases. Semantics-based approaches \cite{7886945, 10.5555/2486788.2486890} take a different route, formulating repair constraints derived from test-suite specifications and solving these constraints to generate patches. Despite their effectiveness, both search-based and semantics-based APR methods often face challenges in scalability and limited patch diversity.

\fakeparagraph{Learning-based APR Techniques}
To address these limitations, learning-based APR techniques emerged \cite{10.1145/3395363.3397369, DBLP:journals/corr/abs-2106-08253}. Early works in this area trained neural machine translation models to predict code fixes, leveraging large corpora from code repositories and incorporating code syntax and semantics. Some studies further improved repair effectiveness by utilizing GitHub issues \cite{Koyuncu_2019} and bug reports as additional training signals. 

\fakeparagraph{LLM-based APR for Repository-level Repair}
Motivated by the need for repository-scale solutions, recent research has turned to LLM-driven, agent-based approaches for automated program repair at the repository level. The advent of benchmarks such as SWE-Bench and its successors \cite{jimenez2024swebenchlanguagemodelsresolve, rashid2025swepolybenchmultilanguagebenchmarkrepository, zan2025multiswebenchmultilingualbenchmarkissue, mathai2024kgym} has enabled systematic evaluation of these techniques by introducing realistic repository-level tasks and reliable test-based validation. These benchmarks have catalyzed extensive research into agent-based frameworks that can autonomously generate and validate patches for complex, multi-file software projects.
The landscape of agent-based repository-level APR can be broadly divided into two main approaches: \textit{(1) Open Process Agent-based Frameworks:} In these frameworks, LLMs act as agents equipped with tools to interact with the software environment~\cite{Cognition, wang2025openhandsopenplatformai, yang2024sweagentagentcomputerinterfacesenable}. The agent dynamically plans and executes actions—such as searching, editing, and testing—based on ongoing feedback, without being restricted to a fixed workflow. Examples include SWE-Agent~\cite{yang2024sweagentagentcomputerinterfacesenable}, which provides interfaces for code navigation and execution, and OpenHands~\cite{wang2025openhandsopenplatformai}, which leverages CodeAct~\cite{wang2024executablecodeactionselicit} to enable a wide range of actions, including internet search. \textit{(2). Workflow-based Frameworks}   These approaches follow a predefined Search-Edit-Test pipeline to address repository-level bugs. The process typically involves localizing faulty code, proposing edits, and validating the fixes through regression tests. Representative systems include Agentless~\cite{xia2024agentlessdemystifyingllmbasedsoftware}, which systematically guides LLMs through the repository to identify context and validate patches; AutoCodeRover~\cite{zhang2024autocoderoverautonomousprogramimprovement}, which integrates program analysis for precise context extraction; SpecRover~\cite{ruan2024specrovercodeintentextraction}, which adds specification generation and patch review; and Patch Pilot~\cite{li2025patchpilot}, which incorporates further patch refinement.

%% file: samples/3_motivation.tex
\section{Motivation}
\label{sec:motivation}

% In this section, we (i) briefly introduce a motivating example from SWE Bench and showcase the common patches observed from other SOTA agents (\S\ref{sec:astropy_issue}), and (ii) explain the need for a deeper understanding of execution (\S\ref{sec:execution_semantics}), issue (\S\ref{sec:issue_semantics}) and code semantics (\S\ref{sec:code_semantics}).  

In this section, we (i) present a motivating example from SWE-Bench and highlight common patches from other SOTA agents (\S\ref{sec:astropy_issue}), and (ii) discuss the need for a deeper understanding of issue (\S\ref{sec:issue_semantics}), and code context (\S\ref{sec:code_semantics}).

\begin{figure*}[pt]
    \centering
    \includegraphics[width=0.66\textwidth]{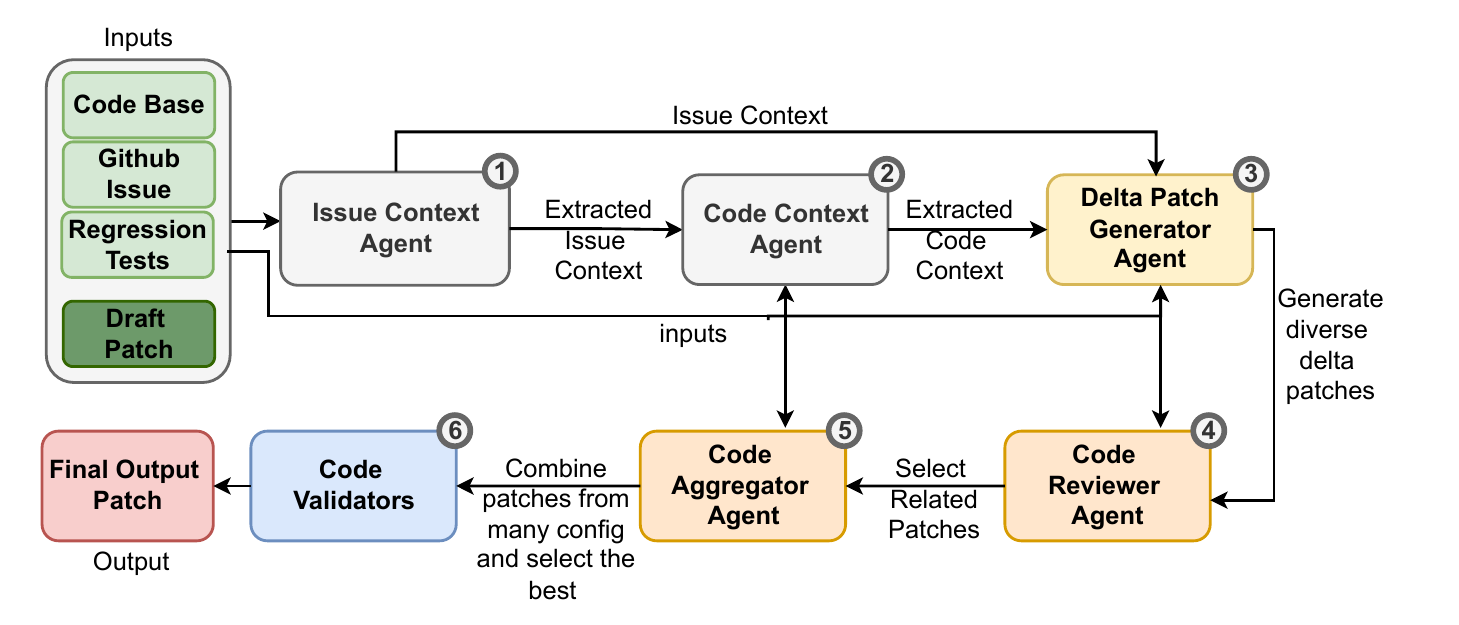}
    \caption[]{\textbf{Pipeline of \ToolName{}}. {\small 
    \ToolName{} takes as input a {codebase}, {GitHub issue}, {regression tests}, and an initial {draft patch}. First, an issue context extractor analyzes the issue description and codebase to extract relevant context \circled{1}. Then, a code context extractor uses this context and the draft patch to select relevant code regions \circled{2}. Leveraging both contexts and the original inputs, \ToolName{} generates diverse delta patches \circled{3} via test-time scaling. A reviewer agent merges each delta with the draft patch \circled{4}, and an aggregator agent ranks them based on alignment with the issue context \circled{5}. Finally, a validation phase \circled{6}—combining test execution and LLM-based judgment—selects the final patch.
    % It takes as input a \textit{codebase}, \textit{GitHub issue}, \textit{regression tests}, and an initial \textit{draft patch}.
    % First, from the issue description and codebase, an issue context extractor extracts issue context \circled{1}. 
    % Next, from the draft patch and issue context, a code context extractor \circled{2} selects the relevant context from the codebase. 
    % Next, leveraging \circled{1} and \circled{2} and the initial inputs \ToolName{} generates generates a delta patch  \circled{3}.
    % We generate diverse delta patches using the principle of test-time scaling.  
    % Next, a reviewer agent combine the delta patches with the draft patch \circled{4} and then aggregator agent \circled{5} selects the most related patches according to the issue context. Finally a validation phase \circled{6} (test run and LLM as judge) chooses the final patch.
    }}
    \label{fig:pipeline}
\end{figure*}

\subsection{Astropy Issue}
\label{sec:astropy_issue}

% To motivate our methodology, we consider astropy-$14635$ -- a sample SWE-Bench Lite problem that most SOTA approaches struggle to solve. As shown in Figure \ref{fig:motivating}, the user-provided issue description (box \blackcircled{1}) details an issue in the Astropy \cite{astropy} library. Concretely, it points out that a specific read functionality for Quick and Dandy Plot (QDP) files does not handle lowercase lines -- expecting everything in uppercase, even though QDP is designed to be case-insensitive. The issue also details a specific example input that causes the library to crash - in this case, the user points out that the string ``read serr 1 2" should be considered a valid input, and that users should not be forced to make this upper-case (i.e. ``READ SERR 1 2"). 

To motivate our approach, we examine \texttt{astropy-14635}, a SWE-Bench Lite task that challenges many SOTA methods. As shown in Figure~\ref{fig:motivating}, the user-reported issue (box \blackcircled{1}) highlights a bug in the Astropy~\cite{astropy} library’s QDP file reader, which incorrectly assumes all input must be uppercase. The report includes a crashing example—``read serr 1 2''—and argues that the reader should accept lowercase input, as QDP is case-insensitive.  We observe that SOTA agents often overanalyze the issue description, producing patches that either are incomplete (i.e., fail to generalize to edge cases)  or overfit to the user inputs (Figure~\ref{fig:motivating}, boxes \blackcircled{3} and \blackcircled{4}, respectively). In contrast, \ToolName{} generates a consistent, generalizable patch aligned with the developer-written fix (box \blackcircled{2}) by leveraging execution, issue, and code semantics.

\subsection{Issue Context}
\label{sec:issue_semantics}

The patch in box \blackcircled{4} illustrates \textit{overfitting}, where the agent fixes only the specific example input in the issue report while missing the broader underlying bug. We find this behavior common in SOTA agents, including Agentless~\cite{xia2024agentlessdemystifyingllmbasedsoftware}, which often hyper-localize based on the issue text. This leads to superficial fixes that act as band-aids rather than addressing root causes. We hypothesize that supplying agents with a deeper understanding of the issue (i.e., \textit{issue semantics}) can mitigate overfitting. Box \blackcircled{5} shows \ToolName{} generated issue semantics, which generalize the problem and explicitly guide edits across multiple components, enabling more robust and complete patches.

\subsection{Code Context}
\label{sec:code_semantics}

The patch in box \blackcircled{3} illustrates an \textit{incomplete} fix—while it addresses the core issue, it overlooks necessary updates elsewhere in the code (e.g., a related \texttt{if} condition). 
Advanced agents like SpecRover \cite{ruan2024specrovercodeintentextraction} are able to avoid ``overfitted'' patches and instead generate such ``incomplete'' patches. We hypothesize that fine-grained code semantics can help agents reason about these secondary changes. Box \blackcircled{6} shows a semantic trace from the QDP module; the red-highlighted step prompts \ToolName{} to identify and apply additional edits, resulting in a complete and consistent patch.

To this end, we show that \textit{issue}, and \textit{code} context are crucial for generating complete patches. 
We now describe \ToolName{}'s overall %pipeline (\S\ref{sec:overview}) and the 
methodologies behind each component (\S\ref{sec:methodology}).

%% file: samples/4_methodology.tex
\section{Methodology}
\label{sec:methodology}

% \TODO{change module ID to match figure}

% \TODO{our approach includes randomness, how to address randomness in our evaluation}

\input{samples/Method/problem}

\input{samples/Method/overview}

\input{samples/Method/context}

\input{samples/Method/delta}

\input{samples/Method/aggregation}

%% file: samples/Method/problem.tex
\subsection{Problem Formulation}

We formally define the \textit{patch refinement} problem as the task of incrementally improving an initial patch, generated by a program repair tool, toward a final, correct repair.
Consider the standard repository-level program repair setting. Let $x = (\mathcal{I}, \mathcal{D}, \mathcal{T})$, where $x$ denotes a program repair instance, fully characterized by the issue statement $\mathcal{I}$, the repository codebase $\mathcal{D}$, and a publicly available regression test suite $\mathcal{T}$. An existing program repair tool, denoted by $\mathsf{R}_{\text{init}}(\cdot)$, takes $x$ as input and generates an initial (seed) patch: $ \mathcal{P}_{\text{init}} = \mathsf{R}_{\text{init}}(x)$.
% $$ \mathcal{P}_{\text{init}} = \mathsf{R}_{\text{init}}(x)$$
Recall from \secref{sec:motivation}, due to over-approximation or under-approximation, the initial {\em draft patch} may be either overfitted or incomplete. The goal of patch refinement is to provide a general solution for further improving these initial seed patches. 
Formally, the objective of patch refinement is to enhance the quality and correctness of $\mathcal{P}_{\text{init}}$ by applying a refinement function $\mathsf{F}(\cdot)$, resulting in a refined, and more accurate patch: 
$ \mathcal{P}_{\text{final}} = \mathsf{F}(\mathcal{P}_{\text{init}})$.

% $$
% \mathcal{P}_{\text{final}} = \mathsf{F}(\mathcal{P}_{\text{init}})
% $$

% We define \ToolName{} as a two-stage patch generation system that incrementally refines an initial patch toward a final, correct repair. Given an issue statement \(\mathcal{I}\), a Python codebase \(\mathcal{D}\), and a regression test suite \(\mathcal{T}\), the objective is to generate a patch \(p\) such that the updated codebase \(\mathcal{D}'\) passes an augmented test suite \(\mathcal{T}'\), which includes additional tests that reproduce the issue described in \(\mathcal{I}\). This process can be defined as: $\mathcal{F}(\mathcal{I}, \mathcal{D}, \mathcal{T}) \rightarrow p$, 
% where, \(\mathcal{F}\) is implemented as a pipeline of the following stages:

%% file: samples/Method/overview.tex
\subsection{Design Overview}

\figref{fig:pipeline} presents the design overview of \tool. Given a program repair instance $x = (\mathcal{I}, \mathcal{D}, \mathcal{T})$ and an initial patch $\mathcal{P}$ generated by an existing APR tool, \tool produces a refined and more accurate patch as output. Specifically, \tool iteratively perform the following three main steps to refine a draft patch:

\begin{enumerate}[leftmargin=*,topsep=0pt]
    \item \textit{Context Extraction and Regularization.} In the first step, (\circled{1} \& \circled{2} of \figref{fig:pipeline}), \tool leverages an LLM agent to extract and regularize context information from both the issue statement and the draft patch. This process mitigates the inherent vagueness and ambiguity often found in issue statements, while also providing a more structured and logical interpretation of the draft patch. As a result, subsequent steps benefit from clearer and more comprehensive contextual information (see \secref{sec:context}).
    
    \item \textit{Diverse Delta Patch Generation}. After extracting the structural context from both the original issue statement and the draft patch, we integrate this information with the draft patch itself and employ an LLM agent to generate multiple diverse delta patches (\circled{3}). In this step, we adopt the philosophy of test-time scaling and perform multiple sampling rounds, enabling us to explore a wider range of possible refinements and ultimately produce a set of diverse delta patches.

    \item \textit{Aggregated Patch Synthesis}. Finally, \tool selects a subset of the generated diverse delta patches, aggregates them with the draft patch, and produces a final refined patch. If the resulting refined patch does not pass the public regression tests, it is treated as the new draft patch for the next iteration (\circled{4}-\circled{6}).

\end{enumerate}

%% file: samples/Method/context.tex
\subsection{Context Extraction and Regularization}
\label{sec:context}

The issue statements collected from GitHub repositories are often vague and ambiguous, as developers frequently use informal or incomplete descriptions. Additionally, draft patches generated by existing APR tools may lack a structured and logical interpretation of the underlying code semantics. To address these limitations and provide more precise context for the LLM agent to reason about the repository, \tool first extracts and regularizes the context from both the original issue statement and the initial draft patch.

Specifically, we begin by defining the context format for both the issue statement and the initial draft patch. We then describe how \tool leverage a LLM agent to extract and regularize this context to support effective downstream reasoning.

\subsubsection{Issue Context Format} 
GitHub issues often include natural language descriptions, examples, and reproduction steps, but LLMs may overfit to specific instances (e.g., block \blackcircled{4} in Figure~\ref{fig:motivating}) or fail to generalize from under-specified inputs. To mitigate this, we compute the \emph{issue context} \(\mathcal{I}' = \mathcal{G}(\mathcal{I}, \mathcal{D})\) by grounding the issue statement \(\mathcal{I}\) in the structure and behavior of the codebase \(\mathcal{D}\). This gives us a structured abstraction that captures the relevant intent, scope, and behavioral expectations underlying a reported software issue. It transforms unstructured natural language into actionable guidance for reasoning about and constructing valid repairs.

Formally, we represent an issue context \(I\) as a 5-tuple: $I = (T, L, A, C, G)$, where:
\begin{itemize}[leftmargin=1.5em]
    \item \( T \): {Target} — The system components implicated by the issue.
    \item \( L \): {Logic} — The intended change in system behavior.
    \item \( A \): {Actions} — High-level plans required to realize the change.
    \item \( C \): {Constraints} — Conditions that must be preserved after the fix.
    \item \( G \): {Generalization} — The scope of applicability beyond the specific instance.
\end{itemize}

For the issue illustrated in Figure~\ref{fig:motivating}, the corresponding issue context is as follows: the \textbf{target} (\(T\)) is the QDP parser’s command matching logic; the \textbf{logic} (\(L\)) involves transforming case-sensitive command recognition into case-insensitive behavior; the \textbf{actions} (\(A\)) include identifying all locations where command strings are compared and updating them to use case-insensitive methods (e.g., normalization or tolerant matching); the \textbf{constraints} (\(C\)) require preserving the parser’s existing behavior for uppercase inputs and avoiding crashes on lowercase inputs; and the \textbf{generalization} (\(G\)) specifies that the fix should apply to all QDP commands, not just the specific instance of \texttt{READ SERR}.

The issue context acts as an intermediate reasoning layer between informal user-reported issues and concrete code-level repair actions. By providing a structured abstraction over the problem space, it guides patch generation in a semantically meaningful and generalizable manner.

\subsubsection{Code Context Format}
%\textbf{Generation of Code Semantics.} 
To generate robust patches, it is not enough to understand the issue alone—\ToolName{} must also capture how the code behaves in relation to the issue and the proposed fix. We define the \emph{code context} as \(\mathcal{C}' = \mathcal{S}(\mathcal{D}, p_{\text{init}})\), which models the interaction between the initial patch and the codebase \(\mathcal{D}\), including call graphs, data flow, control flow, and runtime side effects. This step builds a understanding of code intent, usage, and its relevance to the bug.

To this end, we define \textbf{Patch Context} as a structured representation of the semantic environment surrounding a candidate patch. It captures the relevant data dependencies, control structure, behavioral constraints, and dynamic call relationships that determine how the patch interacts with its surroundings.
%whether a patch is locally correct and safely integrable into the broader program.

We define the patch context \( \mathcal{C}(P) \) for a patch \( P \) at location \( \ell \) as:
$\mathcal{C}(P) = (\mathsf{DD}_\ell, \mathsf{CD}_\ell, \mathsf{IC}_\ell, \mathsf{CG}_\ell)$, where :
\begin{itemize}[leftmargin=1.5em]
    \item \( \mathsf{DD}_\ell \): \textbf{Data Dependencies} — Variables and expressions read/written at \( \ell \), including transitive flows (e.g., line content, regex matches).
    \item \( \mathsf{CD}_\ell \): \textbf{Control Dependencies} — Control-flow constructs affecting the execution of \( \ell \), such as conditional branches and loops.
    \item \( \mathsf{IC}_\ell \): \textbf{Invariant Constraints} — Semantic assumptions and correctness conditions that must be preserved before and after the patch (e.g., input format validity, crash avoidance).
    \item \( \mathsf{CG}_\ell \): \textbf{Call Graph Context} — A backward-traceable call chain showing which higher-level components transitively reach \( \ell \), capturing its functional embedding in the system.
\end{itemize}

For example, in the QDP parser issue described in Figure~\ref{fig:motivating}, the candidate patch modifies the `\_line\_type' function responsible for identifying command lines. The data dependencies include the regular expression `\_command\_re', which encodes the expected command syntax, and the input lines being matched against it. The control dependencies span the call chain from `Table.read', through `QDP.read', etc. % `_read_table_qdp`, `_get_tables_from_qdp_file`, and `_get_type_from_list_of_lines`, down to `_line_type`. 
The invariant constraints require that lowercase commands such as \texttt{read serr 1 2} be recognized as valid without breaking compatibility with existing uppercase-only behavior, and without introducing regressions or crashes elsewhere in the parsing process.

The patch context serves as a semantic scaffold for reasoning about the correctness and consistency of candidate patches, enabling principled refinement and post-hoc validation of learned edits.

\subsubsection{Context Extraction} To ensure that the LLM agent extracts context from the original issue statement and the initial draft patch in accordance with the specified format, we leverage the few-shot learning capabilities of LLMs~\cite{brown2020language} and design two specialized context extraction agents using demonstration examples.
Specifically, we define the extraction process as follows:

$$
\small
\mathcal{I}' = \text{Issue\_Context\_ Extractor}(\mathcal{I}),\quad  \mathcal{C}' = \text{Code\_Context\_Extractor}(\mathcal{P}_{\text{init}})
$$

% \begin{multline}
%   \mathcal{I}' = \text{Issue\_Context\_ Extractor}(\mathcal{I}) \\
%   \mathcal{C}' = \text{Code\_Context\_Extractor}(\mathcal{P}_{\text{init}})
% \end{multline}
where $\mathcal{I}'$ denotes the regularized context extracted from the issue statement, and $\mathcal{C}'$ represents the structured code context extracted from the initial draft patch.

% To refine \(p_{\text{init}}\), \ToolName{} constructs two forms of context based on both the issue and the patched code:

% \begin{itemize}[leftmargin=*]
%     \item \textbf{Issue context:} \(\mathcal{I}' = \mathcal{G}(\mathcal{I}, \mathcal{D})\), computed by grounding the natural language issue statement \(\mathcal{I}\) in the structure and behavior of the codebase \(\mathcal{D}\). This may include extracting relevant call paths, stack traces, involved files and functions, and test artifacts that reproduce the failure.
    
%     \item \textbf{Code context:} \(\mathcal{C}' = \mathcal{S}(\mathcal{D}, p_{\text{init}})\), capturing how the initial patch interacts with  \(\mathcal{D}\), including call graph, data flow, control flow, and runtime side effects.
% \end{itemize}

% Using these contexts, the refinement module generates a set of refined candidate patches:
% \[
% \mathcal{P}' = \mathcal{R}(\mathcal{D}, p_{\text{init}}, \mathcal{I}', \mathcal{C}').
% \]
% Each \(p'_i \in \mathcal{P}'\) is an incremental modification of the initial patch \(p_{\text{init}}\), aimed at resolving semantic gaps and aligning the behavior of the codebase with the intended fix.

%% file: samples/Method/delta.tex
\subsection{Diverse Delta Patch Generation}
\label{sec:patch_gen}

After collecting the regularized contexts, we provide these information to the LLM agent to facilitate deeper reasoning about the limitations of the current initial seed patches and to generate delta patches that enhance the seed patch. To more thoroughly explore the solution space at this stage, we adopt the philosophy of \textit{test-time scaling}, querying the LLM agent multiple times with the same prompt in sampling mode to collect diverse responses. \textit{Test-time scaling} refers to the practice of generating multiple outputs from an LLM during inference—by varying sampling parameters such as temperature—to increase diversity and coverage in the generated solutions, a technique widely studied in prior work \cite{chen2021evaluatinglargelanguagemodels}.

In detail, \tool combines the regularized context with the initial seed patch to construct a prompt, then queries the LLM agent multiple times using this prompt. At each query, the agent is asked to generate an edit in a simple diff format, efficiently producing candidate patches. The diff format includes two parts: the search snippet (the code to be replaced) and the replacement snippet (the new code to insert).
To apply the generated diff to the original patch, we simply match the search code snippet in the codebase and replace it with the replacement. This simple diff format avoids generating the complete code and instead focuses on producing small, targeted edits, which are not only more cost-efficient but also more reliable and accurate, with reduced risk of hallucinations.
These delta patch variants reflect diverse perspectives on the issue and repair process. An aggregator module then reconciles the resulting set of refined patches into a unified, semantically consistent pool that preserves both the intent of the issue and the correctness of the code.

% \textit{Test-Time Scaling for Robust Patch Selection.}   To further enhance the robustness of patch refinement, we adopt a \emph{test-time scaling} strategy~\rayb{cite}. At inference time, we collect candidate patches from multiple points in the refinement pipeline—each informed by different semantic configurations (e.g., with or without refinement or reviewer stages; see RQ2). These variants reflect diverse perspectives on the issue and repair process. An aggregator then reconciles the resulting set of refined patches into a unified, semantically consistent pool that preserves both the issue intent and the correctness of the code. This pool is evaluated by an LLM-based voting agent (\circled{15}), configured with a high temperature to encourage exploratory scoring across candidates. The patch with the highest number of votes over multiple sampling rounds is selected as the final output. This test-time scaling mechanism amplifies the strengths of diverse repair strategies and improves the likelihood of selecting a robust, generalizable, and developer-aligned fix.

%% file: samples/Method/aggregation.tex
\subsection{Aggregated Patch Synthesis}

After collecting a set of diverse delta patches from \secref{sec:patch_gen}, the next step is to aggregate these patches into a single, valid patch that can effectively address the bugs described in the original issue statement.
To achieve this, \tool incorporates three steps: a \textit{code reviewer agent}, an \textit{aggregation agent}, and an \textit{patch validator}

\subsubsection{Code reviewer agent} While \textit{test-time scaling} helps explore a broader search space and enables the generation of more diverse patches, it can also produce unrelated or irrelevant patches. To address this, \tool simulates the code review process by leveraging the concept of LLM-as-Judge and employs a dedicated \textit{code reviewer agent}. Specifically, \tool combines each delta patch with the seed patch and the issue context, then queries the LLM agent to determine whether the revised patch adequately addresses the issue described in the context. The LLM agent is asked to provide a simple yes or no response. In this way, \tool effectively filters out unrelated patches, retaining only those that are relevant to the issue.

\subsubsection{Aggregation agent} After filtering out unrelated delta patches, \tool first performs de-duplication to remove redundant patches. {Next, it groups the remaining delta patches based on conflicts-i.e., when two patches modify the same line differently, similar to a merge conflict. To do this, all patches are passed to the LLM, which is prompted to group conflicting ones; all patches in a group along with the issue description, are then passed to the LLM for aggregation. Here aggregation works like a developer resolving a merge conflict: if patches are subsets, their changes are merged; if they diverge, the LLM selects or combines changes based on the issue description, producing one unified delta patch per group. These non-conflicting unified patches are then merged with the initial draft patch using an LLM that checks for overlaps or conflicts and produces a consolidated set of diffs.

These diffs, containing the old and new code snippets, are passed to a git-diff extraction program that locates the original code in the codebase and replaces it with the new code while handling indentation mismatches, and finally the git diff command is run to generate the final consolidated patch. We also ensure syntactic correctness during aggregation by linting the code (e.g., with pylint) to catch syntax errors whereas semantic correctness is addressed iteratively - if a patch introduces a semantic issue, it is corrected in the next iteration when that patch serves as the draft.

\subsubsection{Patch Validator}
After generating the consolidated final patch, we validate its correctness using the publicly available regression tests. If the final patch passes all regression tests, \tool returns it as the solution. Otherwise, the final patch is treated as a new initial seed patch, and the process is repeated until the maximum number of retry iterations is reached. Our evaluation results (~\secref{sec:res_rq3}) show that even with just one retry iteration, \tool{} demonstrates significant improvements in bug-fixing accuracy.

%% file: samples/5_experiments.tex
\section{Experimental Setup}
\label{sec:setup}

In this paper, we conduct empirical evaluations and seek to answer the following research questions, due to the space limit, more results could be found on our website.

\begin{enumerate}[label=\textbf{RQ\arabic*}.]

\item \label{rq:rq2} \textbf{Plugin Effectiveness:}
To what extent does integrating \tool as a plugin enhance the performance of different APR approaches?

\item \label{rq:rq1} \textbf{Bug-Fixing Capability:} 
What is the overall performance of \tool in resolving real-world bugs and generating valid patches?

\item \label{rq:rq3} \textbf{Module Contribution:}
What is the contribution of each individual module within \tool to its overall performance and effectiveness?

    % \item \label{rq:sensitivity} 
    %  \textbf{Stability:}
    % Can \ToolName{} consistently produce reliable and correct patches across diverse settings?
\end{enumerate}

\input{samples/setup/dataset}
\input{samples/setup/baseline}

\input{samples/setup/metrics}

\input{samples/setup/process}

\input{samples/setup/implementation}

%% file: samples/setup/dataset.tex
\subsection{Evaluation Dataset}
We evaluate \ToolName{} on SWEBench Lite \& Verified ~\cite{jimenez2024swebenchlanguagemodelsresolve}, benchmarks consisting of 300 \& 500 real-world GitHub issues across 12 Python repositories. For each issue, \ToolName{} receives the issue description and pre-fix codebase with regression tests, and outputs a single patch in git diff format, which is evaluated for successful resolution.

%% file: samples/setup/baseline.tex
\subsection{Comparison Baselines}

To evaluate the performance of \tool, we compare it with a comprehensive range of state-of-the-art approaches in Automated Program Repair (APR). The selected baselines encompass two primary methodologies: manually defined workflow-based frameworks and more autonomous agentic frameworks. This comparison allows us to contextualize the contributions of \tool and demonstrate its efficacy.

\fakeparagraph{Workflow-Based Approaches}
These methods adopt a structured search–edit–test sequence, with dedicated modules orchestrating each stage.
\begin{itemize}[leftmargin=*]
    \item \textbf{\texttt{AutoCodeRover}~\cite{zhang2024autocoderover}:} A repair pipeline combining AST-aware code search, spectrum-based fault localization, and iterative patch generation.
    \item \textbf{\texttt{Agentless-1.5}~\cite{xia2024agentlessdemystifyingllmbasedsoftware}:} Emphasizes modular and sequential processing of search, edit, and test stages without integrated agentic reasoning.
    \item \textbf{\texttt{SpecRover}~\cite{ruan2024specrovercodeintentextraction}:} Improves upon AutoCodeRover by using developer intent signals for enhanced fault localization.
    \item \textbf{\texttt{ExpeRepair-v1.0}~\cite{li2025experepair}:} Leverages past repair experiences and examples to generate more effective patches.
    \item \textbf{\texttt{OrcaLoca + Agentless-1.5}~\cite{liu2024orcaloca, xia2024agentlessdemystifyingllmbasedsoftware}:} Extends the \texttt{Agentless} framework with the \texttt{OrcaLoca} module for improved fault localization.
\end{itemize}

\fakeparagraph{Agent-Based Approaches}
These tools grant greater autonomy to the underlying models, allowing them to interact more dynamically with the development environment.
\begin{itemize}[leftmargin=*]
    \item \textbf{\texttt{SWE-agent}~\cite{yang2024sweagentagentcomputerinterfacesenable}:} Employs a ReAct-style loop to iteratively interact with a sandboxed coding environment.
    \item \textbf{\texttt{OpenHands}~\cite{wang2025openhandsopenplatformai}:} A generalist agent framework that executes shell commands and modifies codebases to complete tasks.
    \item \textbf{\texttt{Moatless Tools}~\cite{ghissassi2024moatless}:} Applies Monte Carlo Tree Search (MCTS) to systematically explore the solution space.
    \item \textbf{\texttt{DARS Agent}~\cite{aggarwal2025dars}:} Utilizes Dynamic Action Re-Sampling to recover from suboptimal decisions by exploring alternative actions.
    \item \textbf{\texttt{devlo}~\cite{devlo2024}:} An AI developer agent designed to autonomously resolve GitHub issues.
    \item \textbf{\texttt{Blackbox AI Agent}~\cite{blackboxai2024}:} A closed-source agent for automated coding assistance and bug-fixing.
    \item \textbf{\texttt{Globant Code Fixer Agent}~\cite{globant2024codefixer}:} An autonomous agent designed to identify, analyze, and automatically fix bugs.
    \item \textbf{\texttt{CodeV}~\cite{li2024codev}:} A multi-agent framework designed for collaborative code generation and repair.
    \item \textbf{\texttt{Codart AI}~\cite{codart2024}:} Analyzes code context to generate precise, targeted patches for identified bugs.
    \item \textbf{\texttt{CodeStory Aide}~\cite{amann2024codestory}:} Leverages the history and context of code changes to inform the repair process.
    \item \textbf{\texttt{Lingxi}~\cite{yang2025lingxi}:} An open-source, multi-agent framework that coordinates specialized agents within a graph-based workflow.
\end{itemize}

For all baseline approaches, we follow the protocol of prior work~\cite{xia2024agentlessdemystifyingllmbasedsoftware} and report their best performance metrics as published in their respective papers, official websites, blogs, or leaderboard and the seed patches produced by these APR approaches, which we use in our experiments, can be obtained from the SWEBench website \cite{swebenchwebsite}.

%% file: samples/setup/metrics.tex
\subsection{Evaluation Metrics}

\fakeparagraph{Correctness Metrics} 
To evaluate the correctness of our method in addressing real-world bugs, we adopt the \textit{Resolved Issue Rate}~\cite{zhang2024autocoderover} as our evaluation metric, following prior work. The \textit{Resolved Issue Rate} is defined as the percentage of generated patches that successfully pass all hidden test cases.

\fakeparagraph{Bug Localization Metrics} 
In addition to the \textit{Resolved Issue Rate}, we also report the \textit{Correct Location Rate} in accordance with established practice. The \textit{Correct Location Rate} measures the percentage of problems for which the patch generated by the tool includes the edit locations present in the ground-truth developer patch. Consistent with previous work, we evaluate this metric at file granularity. A patch is considered to contain the correct location if it modifies a superset of all the locations specified in the ground-truth patch. 

\fakeparagraph{Cost Metrics}
Besides the correctness evaluation, we also consider two cost metrics. The first metric is \textit{Average Cost}, which measures the average monetary cost of running the tool. The second metric is \textit{Average Tokens}, for which we report both the input and output tokens required by our tool to generate a patch for one GitHub issue.

%% file: samples/setup/process.tex
\subsection{Experiment Process}

\fakeparagraph{\ref{rq:rq2} Process} To address this question, we apply \tool on a diverse set of state-of-the-art APR approaches including AutoCodeRover, Agentless, CodeV, BlackBoxAI, and ExpeRepair - spanning closed and open-source systems, workflow and agentic-based methods and different LLMs. We then use \tool to improve a subset of thirty initial patches generated by these methods and measure the resolution rate both with and without \tool. These thirty issues were selected by repeatedly sampling from the full distribution until we found a subset where AutoCodeRover resolved similar number of issues in the subset (36.67\%) vs the full dataset (37\%), it had at least one issue from each repository, and the resulting distribution was similar to the full dataset using a Chi-square goodness-of-fit test \& ensuring that the p-value exceeded 0.05 (obtained p-value was 0.0935). By comparing the results before and after applying \tool{}, we can assess its effectiveness as a general-purpose plugin for enhancing current APR tools. Secondly, we perform a fine-grained analysis to identify the number of unique issues that are successfully fixed by \tool using AutoCodeRover as the seed, but were not solved by any other SOTA on the full SWEBench dataset. 

\fakeparagraph{\ref{rq:rq1} Process} To address this research question, we evaluate the overall effectiveness of \tool on the SWEBench benchmark when using AutoCodeRover as the seed and compare it to the performance of other approaches. For each issue, we use \tool to generate a patch then assess its correctness using the hidden test cases, following the SWE-Bench standards. A patch is considered correct only if it passes all hidden test cases.

\fakeparagraph{\ref{rq:rq3} Process} To answer this question, we conduct several experiments to evaluate the contribution of each module within \tool{} and to investigate how its hyperparameters affect overall performance. First, we perform ablation studies by removing each core module from \tool—specifically, the \textit{context regularization}, \textit{diverse delta patch generation}, and \textit{code reviewer} modules—and then measure the bug-fixing capability of the modified system. 
Second, we examine the impact of different backend LLMs on the overall performance of \tool. We evaluate the system using single backend models, such as \texttt{Claude3.7-Sonnet} and \texttt{Gemini2.5-Pro}, and further explore how varying the backend LLM specifically for the \textit{code reviewer} module affects \tool's effectiveness. 
Finally, we assess \tool's performance under various hyperparameter settings to understand their influence on the results. In particular, we investigate the \textit{number of retry loops} used for iterative patch refinement, which balances the trade-off between cost and search space, as well as the \textit{temperature} parameter, which controls the randomness and diversity of the LLM’s output.

%% file: samples/setup/implementation.tex
\subsection{Implementation Details}

\tool{} requires an existing APR tool to generate the initial patch, which serves as the seed for further refinement. In our experiments on SWEBench Verified \& Lite \secref{sec:res_rq_results}, we use AutoCodeRover\cite{zhang2024autocoderover} as the default underlying APR approach for seed patch generation, due to its strong performance, extensibility and open-source nature. In \secref{sec:res_rq2}, we replace the seed generation module with other existing APR tools to investigate how \tool can enhance different APR approaches. For AutoCodeRover configuration, we set the number of agent conversation rounds during localization and API extraction to 15. Empirically, fewer than 8 rounds often result in incomplete localization. Since the localization process can terminate early if the LLM confirms that the correct file has been localized, this cap helps avoid unnecessary LLM calls.

\tool is implemented using \texttt{Claude 3.7 Sonnet} as the backend LLM and \texttt{Gemini 2.5 Pro} as the reviewer agent for delta patch aggregation. In \secref{sec:res_rq3}, we further evaluate the impact of different backend LLMs on the overall performance of \tool{}.

By default, \tool{} employs greedy sampling with a temperature set to zero for deterministic results during delta patch generation. For more diverse generation, we set the temperature to 0.7 and run up to 5 retry loops per issue, terminating early if a suitable patch is found. We evaluate the impact of both the \textit{temperature} and the \textit{number of retry loops} in \secref{sec:res_rq3} to understand how these hyperparameters influence the performance of \tool.

To evaluate the correctness of each APR tool in fixing real-world GitHub issues, we strictly follow the SWE-bench benchmark guidelines. In this setting, the inputs include a user-submitted issue description (typically in natural language), the complete codebase of the repository, and a set of public test cases (usually from regression tests). The goal is to automatically generate a correct patch by reasoning over the entire codebase. The correctness of each generated patch is then assessed using private, rigorous test suites to ensure comprehensive validation. In accordance with the SWE-bench protocol, each tool is allowed a single \textit{pass\@1} attempt, without access to test-specific metadata, hints, or external web resources. All generated patches must be self-contained and executable within the provided context.

% \TODO{not sure whether need to describe each module here or in appraoch}

%% file: samples/6_results.tex
\section{Results}
\label{sec:results}

\input{samples/Results/rq2}
\input{samples/Results/rq1}

\input{samples/Results/rq3}

%% file: samples/Results/rq2.tex
\input{samples/Table/diff_base}

\subsection{\ref{rq:rq2} Results}

\label{sec:res_rq2}

\fakeparagraph{Overall Enhancement} The overall effectiveness of \tool{} in enhancing existing APR tools is presented in \tabref{tab:diff_base}. In this table, the ``APR'' column lists the baseline APR tool names, ``Initial'' shows the issue resolution rate achieved by the original tools, and ``Post refinement'' reports the resolution rate after applying \tool{}’s refinement process. The “Increase” column quantifies the improvement brought by \tool{}. From the results, we observe that integrating \tool{} leads to consistent improvements across all baseline tools, with resolution rates increasing by up to 20\% and 14\% on average. This demonstrates \tool{}’s ability to robustly enhance the bug-fixing performance of various APR methods, regardless of their original effectiveness.

\begin{figure}
    \centering
    \includegraphics[width=0.39\textwidth]{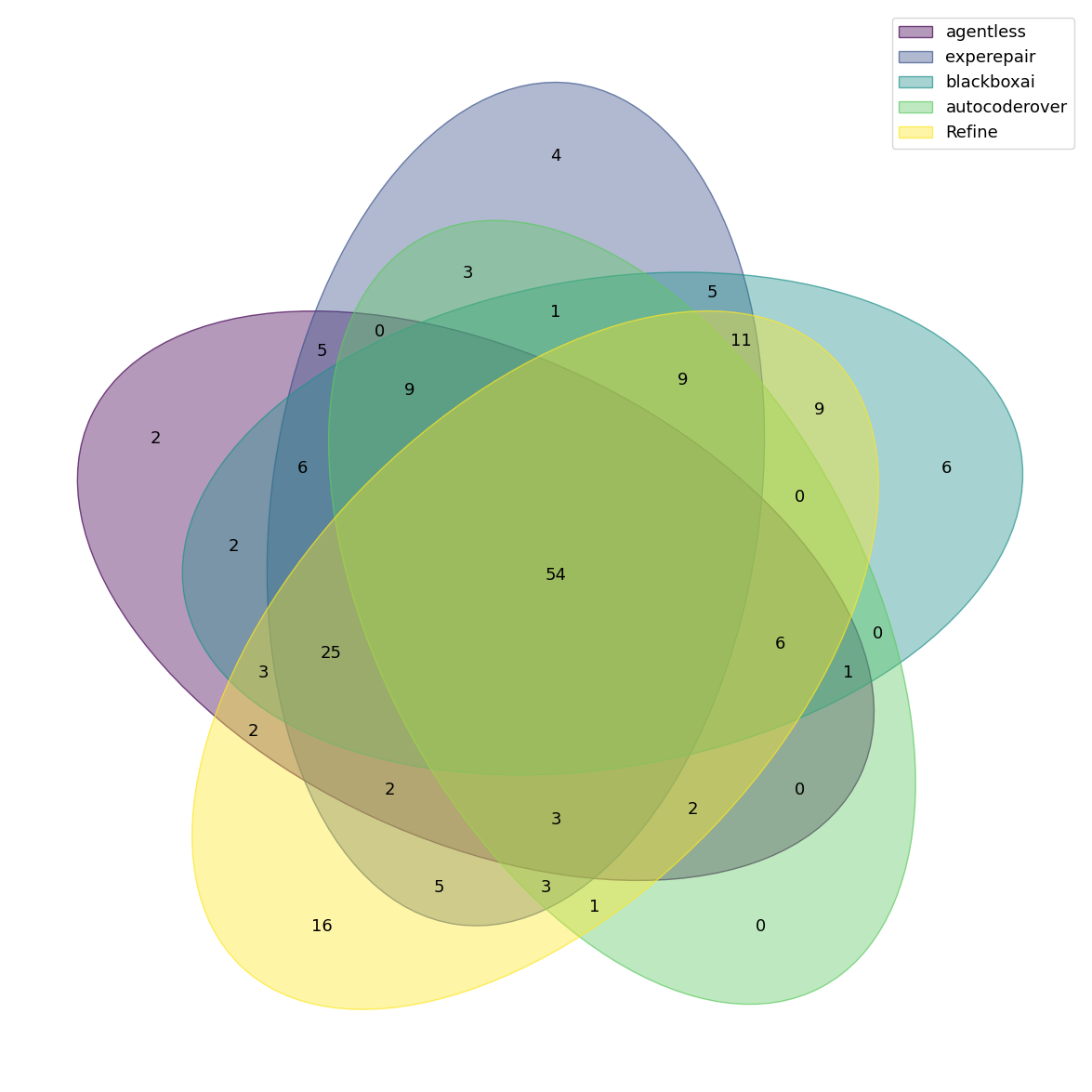}
    \caption{Fine-grained patch analysis: Venn diagram comparing the sets of issues resolved by Five APR approaches on SWEBench. \tool{} (built on top of \textsc{AutoCodeRover}) resolves the largest number of unique issues (16)}
    \label{fig:patch_analysis}
\end{figure}
\fakeparagraph{Fine-grained Fix Analysis} \figref{fig:patch_analysis} presents a fine-grained comparison of fixed issues across different APR approaches. The Venn diagram illustrates overlap and uniqueness in resolved issues among five methods: four baseline APR tools as mentioned above based on their official SWEBench submissions, and \tool{} when refining patches generated by \textsc{AutoCodeRover}. Notably, \tool{} resolves 16 unique issues not fixed by any other method while also outperforming all other baselines, each of which resolves at most 6 unique issues. An interesting observation is that different base APR tools produce notably distinct sets of patches, with substantial variation in the bugs each tool can resolve. The overlaps between the sets indicate that while there is some commonality in the bugs fixed, each APR approach also contributes uniquely, capturing bugs that others may miss. Notably, the integration of our refinement module (\textsc{Refine}) enables the recovery of a significant number of additional unique patches, further improving the overall coverage. These results yield two key takeaways. First, patch refinement is a highly effective strategy for boosting APR performance, not only by significantly increasing resolution rates (as shown in \tabref{tab:diff_base}) but also by expanding bug coverage to fix unique issues that other tools miss (\figref{fig:patch_analysis}). Second, the success of \tool{}'s refinement process highlights the critical importance of leveraging both high-level issue descriptions and fine-grained code context in resolving difficult multi-function bugs. 

%% file: samples/Table/diff_base.tex
\begin{table}[htbp]
  \centering
  \caption{The overall effectiveness of \tool{} in enhancing existing APR tools performance on SWEBench Lite}
  \resizebox{0.38\textwidth}{!}{
    \begin{NiceTabular}{lrrr}
    \CodeBefore
        \rowcolors{3}{}{blue!8}
        \Body
    \toprule
    \toprule
    \textbf{APR} & \multicolumn{1}{l}{\textbf{Initial}} & \multicolumn{1}{l}{\textbf{Post refinement }} & \multicolumn{1}{l}{\textbf{Increasement}} \\
    \midrule
    \textbf{AutoCodeRover} & 36.67\% & 56.67\% & 20\% \\
    \textbf{Codev} & 50\%  & 53.33\% & 3.33\% \\
    \textbf{ExpeRepair} & 40\%  & 60\% & 20\% \\
    \textbf{Agentless} & 40\%  & 56.67\% & 16.67\% \\
    \textbf{BlackBoxAI} & 50\%  & 60\%  & 10\% \\
    \bottomrule
    \bottomrule
    \end{NiceTabular}%
    }
  \label{tab:diff_base}%
\end{table}%

%% file: samples/Results/rq1.tex
\input{samples/Table/rq1}

\subsection{\ref{rq:rq1} Results}

\label{sec:res_rq_results}

\fakeparagraph{Bug-Fix Correctness}
\tabref{tab:rq1} shows the performance of \tool{} against state-of-the-art APR baselines on SWE-Bench Lite \& Verified. \tool{} successfully resolves 155 issues (51.67\%), outperforming all baselines with a 14.67 percentage point improvement over its initial seed. This increase, which accounts for 44 additional resolved issues, highlights the effectiveness of \tool{} in refining patches leveraging issue and code semantics. We further validate this on SWE-Bench Verified, where \tool{} refines patches from AutoCodeRover to improve the resolution rate from 51.6\% to 63.8\% - with an absolute increase of 12.2 percentage points and correctly resolving an additional 61 issues.

\fakeparagraph{Bug Localization}
The bug localization performance of \ToolName{} is driven by its patch refiner. This component detects when an initial localization is fundamentally incorrect and uses that information to guide a more precise attempt in a subsequent round if applicable.

\fakeparagraph{Cost Analysis} On average, our approach requires 22.4 minutes to solve an issue, with a cost of \$6.59 and 1.15 million tokens per issue under the parameters described in the \secref{sec:setup}. These costs can be reduced to \$4.77 per issue across multiple runs when utilizing issue semantics and repair stage caching. Notably, the median cost per issue is \$4.87, which is substantially lower than the average due to a small number of costly outliers—such as the \texttt{scikit-14087} GitHub repository (over \$30)—that skew the average cost.

%% file: samples/Table/rq1.tex
\begin{table*}[htbp]
  \centering
  \caption{Comparison of Bug-Fixing Accuracy, Cost, and Patch Location Correctness Between \tool and Baseline Methods}
  \resizebox{0.98\textwidth}{!}{%
    \begin{NiceTabular}{llcc ccc}
    \CodeBefore
        \rowcolors{3}{}{blue!8}
        \Body
    \toprule
    \toprule
    \multirow{2}[2]{*}{\textbf{Approach}} & \multirow{2}[2]{*}{\textbf{LLM}} & \multicolumn{2}{c}{\textbf{Correctness}} & \textbf{\% Correct Location} & \multicolumn{2}{c}{\textbf{Cost}} \\
          &       & \textbf{SWE-bench Lite} & \textbf{SWE-bench Verified} & \textbf{File} & \boldmath{}\textbf{Avg. Cost (\$)}\unboldmath{} & \textbf{Avg. Tokens} \\
    \midrule
    \textbf{Moatless Tools} & Claude 3.5 Sonnet & 38.30\% (115) & N/A   & N/A   & 0.17  & N/A \\
    \textbf{Codart AI} & Claude 3.5 Sonnet & 41.67\% (125) & N/A   & N/A   & N/A   & N/A \\
    \textbf{Openhands} & CodeAct v2.1 & 41.67\% (125) & 53.00\% (265)   & N/A   & 2.14  & N/A \\
    \textbf{Lingxi} & N/A   & 42.67\% (126) & N/A   & N/A   & N/A   & N/A \\
    \textbf{CodeStory Aide} & N/A   & 43.00\% (129) & 62.20\% (311)   & N/A   & N/A   & N/A \\
    \textbf{DARS Agent} & Claude 3.5 Sonnet & 47.00\% (141) & N/A   & N/A   & 12.24 & N/A \\
    \textbf{devlo} & Claude 3.5 Sonnet & 47.33\% (142) & N/A   & N/A   & N/A   & N/A \\
    \textbf{SWE-agent} & Claude 3.7 Sonnet & 48.00\% (144) & 62.40\% (312)    & N/A   & 1.62  & 521k \\
    \textbf{Blackbox AI Agent} & Claude 3.5 Sonnet & 49.00\% (147) & 62.80\% (314)   & 75.00\% & N/A   & N/A \\
    \textbf{Globant Code Fixer Agent} & N/A   & 48.33\% (145) & N/A   & N/A   & N/A   & N/A \\
    \textbf{Gru} & N/A   & 48.67\% (146) & 57\% (285)   & N/A   & N/A   & N/A \\
    \textbf{Codev} & Claude 3.5 Haiku + Gemini 2.5 Pro + o4-mini & 49.00\% (147) & N/A   & 79.00\% & N/A   & N/A \\
    \midrule
    \textbf{ExpeRepair-v1.0} & Claude 3.5 Sonnet + o3-mini & 48.33\% (145) & N/A   & 81.00\% & 2.07  & N/A \\
    % \textbf{AutoCodeRover} & GPT-4o & 30.67\% (92) & 51.60\% (258)   & 69.3\% (208) & 0.65  & 37.6k \\
    \textbf{AutoCodeRover} & N/A   & 37.00\% (111) & 51.60\% (258)   & N/A   & N/A   & N/A \\
    \textbf{Agentless-1.5} & Claude 3.5 Sonnet & 40.67\% (122) & 50.80\% (254)   & 76.67\% & 1.12  & N/A \\
    \textbf{OrcaLoca + Agentless-1.5} & Claude 3.5 Sonnet & 41.00\% (123) & N/A   & N/A   & N/A   & N/A \\
    \midrule
    \textbf{\tool} & Claude 3.7 Sonnet + Gemini 2.5 Pro & 51.67\% (155) & 63.8\% (319)   & 85.67\% & 6.59 & 1.15M \\
    \bottomrule
    \bottomrule
    \end{NiceTabular}%
    }
  \label{tab:rq1}%
\end{table*}%

%% file: samples/Results/rq3.tex
\subsection{\ref{rq:rq3} Results}

\label{sec:res_rq3}

\input{samples/Table/ablation}

\fakeparagraph{Ablation Study} \tabref{tab:ablation} presents the results of an ablation study evaluating the impact of each module within our approach on both the resolved issue rate and accuracy drop. The results clearly demonstrate that the inclusion of each module contributes to improved bug-fixing performance. The baseline configuration, which omits all three modules, achieves a resolved issue rate of 37.00\% with an accuracy drop of -7.66\%. Adding any two modules provides only marginal improvement. Notably, the inclusion of the context extraction and code reviewer modules without the diverse delta patch generation module yields a 40.33\% resolved rate and reduces the accuracy drop to -4.33\%. Introducing the diverse delta patch generation module further boosts performance, with the combination of all three modules achieving the highest resolved issue rate of 44.66\% when strictly using Claude 3.7 Sonnet and eliminating the observed accuracy drop. These results highlight the complementary effect of the three modules, particularly the importance of integrating both context extraction and code review for maximizing patch correctness and reliability.

\input{samples/Table/llm}

\fakeparagraph{Backend LLMs Impact} \tabref{tab:llm} summarizes the impact of using different LLM backends in each module of the \tool framework. The baseline configuration, which is the original APR tool without the refinement process (i.e. AutoCodeRover), achieves a resolved issue rate of 37\%. Replacing all three modules with Claude 3.7 Sonnet (C-3.7) significantly increases the resolved issue rate to 44.66\% with an accuracy improvement of 7.66\%. Switching the backend of all modules to Gemini 2.5 Pro (G-2.5) further improves performance, yielding a resolved issue rate of 50.33\% and an accuracy increase of 13.33\%. The best performance is achieved when the context extraction and delta generation modules use C-3.7, while the reviewer module uses G-2.5, resulting in a resolved issue rate of 51.67\% and the highest accuracy gain of 14.67\%. Using C-3.7 for context extraction and delta generation with C-4 or with a combination of multiple reviewers (C-3.7, C-4, and G-2.5) also leads to substantial improvements, although not surpassing the mixed C-3.7 and G-2.5 setup. These results highlight not only the importance of leveraging powerful LLM backends but also the benefits of module-specific backend selection, with a heterogeneous configuration offering the greatest gains in both correctness and accuracy.

\input{samples/Table/hyperparameters}

\fakeparagraph{Hyperparameters Impact} \tabref{tab:hyper} presents the performance of \tool under different hyperparameter settings when using C-3.7 \& G-2.5, specifically varying the number of retry loops and the temperature parameter. The results show that increasing the number of retry loops consistently improves the resolved issue rate: from 49.00\% with 1 retry, to 50.00\% with 3 retries, and reaching 51.67\% with 5 retries (all at a temperature of 0.7). Similarly, varying the temperature parameter while keeping the number of retries fixed at 5 also impacts performance. A temperature of 0.0 yields a resolved issue rate of 48.33\%, while increasing the temperature to 0.3 improves the rate to 49.33\%, and a temperature of 0.7 achieves the highest rate of 51.67\%. These findings indicate that both a higher number of retry loops and an increased temperature setting can enhance the effectiveness of the patch refinement process, likely by promoting greater exploration and diversity in the generated patches.

%% file: samples/Table/ablation.tex
% Table generated by Excel2LaTeX from sheet 'Sheet3'
\begin{table}[htbp]
  \centering
  \caption{Ablation Study: Impact of Each Module on Issue Resolution Rate and Accuracy Drop. "Ctx." = Context Extraction, "Div Delta" = Diverse Delta Patch Generation, "Reviewer" = Code Reviewer.}
  \resizebox{0.38\textwidth}{!}{
    \begin{NiceTabular}{ccc|cc}
    \CodeBefore
        \rowcolors{3}{}{blue!8}
        \Body
    \toprule
    \toprule
    \multicolumn{3}{c|}{Module} & \multirow{2}[2]{*}{Resolved Issue Rate} & \multirow{2}[2]{*}{Acc Drop} \\
    \textbf{Ctx.}  &  \textbf{Div Delta} & \textbf{Reviewer}  &       &  \\
    \midrule
          &       &       & 37.00\%  & -7.66\% \\
          & \checkmark & \checkmark & 37.33\% & -7.33\% \\
    \checkmark &       & \checkmark & 40.33\% & -4.33\% \\
    \checkmark & \checkmark &       & 42.00\%  & -2.66\% \\
    \checkmark & \checkmark & \checkmark & 44.66\% & - \\
    \bottomrule
    \bottomrule
    \end{NiceTabular}%
    }
  \label{tab:ablation}%
\end{table}%

%% file: samples/Table/llm.tex
% Table generated by Excel2LaTeX from sheet 'Sheet3'
\begin{table}[htbp]
  \centering
  \caption{The performance of \tool  with different LLM backend}
  \resizebox{0.36\textwidth}{!}{
    \begin{NiceTabular}{ccc|cc}
    \CodeBefore
        \rowcolors{3}{}{blue!8}
        \Body
    \toprule
    \toprule
    \multicolumn{3}{c|}{Module} & \multirow{2}[2]{*}{Resolved Issue Rate} & \multirow{2}[2]{*}{Acc Inc} \\
    \textbf{Ctx.}  &  \textbf{Div Delta} & \textbf{Reviewer}  &       &  \\
    \midrule
    -     & -     & -     & 37\%  & - \\
    C-3.7  & C-3.7  & C-3.7  & 44.66\% & 7.66\% \\
    C-3.7  & C-3.7  & C-4   & 46.00\% & 9.00\% \\
    C-3.7  & C-3.7  & C-3.7 + C-4 +  G-2.5 & 46.33\% & 9.33\% \\
    G-2.5 & G-2.5 & G-2.5 & 50.33\% & 13.33\% \\
    C-3.7  & C-3.7  & G-2.5 & 51.67\% & 14.67\% \\
    \bottomrule
    \bottomrule
    \end{NiceTabular}%
    }
  \label{tab:llm}%
\end{table}%

%% file: samples/Table/hyperparameters.tex
\begin{table}[htbp]
  \centering
  \caption{Performance of \tool under different hyperparameters}
  \resizebox{0.38\textwidth}{!}{
    \begin{NiceTabular}{ccc}
    \CodeBefore
        \rowcolors{2}{}{blue!8}
        \Body
    \toprule
    \toprule
    \textbf{Hyperparameters} & \textbf{Approach  Setting} & \textbf{Resolved Issue Rate} \\
    \midrule
    \multirow{3}[2]{*}{\textbf{\# of Retry Loops}} & 1 Retry ( Temp 0.7) & 49.00\% (147) \\
          & 3 Retries (Temp 0.7) & 50.00\% (150) \\
          & 5 Retries (Temp 0.7) & 51.67\% (155) \\
    \midrule
    \multirow{3}[2]{*}{\textbf{Temperature}} & Temp 0.0 (5 Retries) & 48.33\% (145) \\
          & Temp 0.3 (5 Retries) & 49.33\% (148) \\
          & Temp 0.7 (5 Retries) & 51.67\% (155) \\
    \bottomrule
    \bottomrule
    \end{NiceTabular}%
    }
  \label{tab:hyper}%
\end{table}%

%% file: samples/7_case_study.tex
\input{samples/CaseStudy/pallets_4045}

%% file: samples/CaseStudy/pallets_4045.tex
\subsection{How \tool{} Resolves Pallets-4045  - An Issue That No Other APR Could Resolve}

Finally, we take a look at how \tool{} refines a partially correct patch from AutoCodeRover to resolve a Flask issue that \textbf{none} of the top SOTA models have been able to successfully fix similar to Astropy 14635 in our motivating example.

\textbf{Issue Overview of Pallets-4045:} The issue arises from Flask’s support for nested Blueprints, where the dot character serves as a delimiter. To avoid ambiguity, Blueprint-related names (e.g., for endpoints, view functions, CLI groups) must not contain dots.

\textbf{From Initial Draft to Complete Solution:}
The refinement process highlights how a basic fix can evolve into a robust solution through contextual reasoning.

\textbf{Initial Patch:} AutoCodeRover's patch introduces a check in the Blueprint constructor:

\begin{lstlisting}[language=diff, 
  basicstyle=\footnotesize,
  breaklines=true,
  frame=single,                 
  framerule=0.5pt,              
  rulecolor=\color{black}, 
  ]
+    if "." in name:
+        raise ValueError("Blueprint names should not contain a dot.")
\end{lstlisting}

While this handles the main issue, it misses other cases—such as CLI groups and view function names—where dots can also cause conflicts. This patch can be thought of as a near miss draft patch since it solves the bulk of the issue but doesn't fully resolve the issue.

\textbf{Refined Patch:} \tool{} identifies relevant code contexts of the Blueprint class such as the call chain of the function or code required for CLI registration or URL routing and uses this code context to make context aware fixes that patch the missing edge cases in conjunction with the issue and initial patch. It then adds validations for CLI group names, replaces brittle \texttt{assert} statements with robust \texttt{ValueError} exceptions, and introduces logic to check view function names only when an endpoint is not specified. This progressive enhancement ensures a comprehensive and correct fix. \\\\

\begin{lstlisting}[language=diff, 
  basicstyle=\footnotesize,
  breaklines=true,
  frame=single,                 
  framerule=0.5pt,              
  rulecolor=\color{black}, 
  ]
+    if "." in name:
+        raise ValueError("Blueprint names should 
...
+    if cli_group is not _sentinel and cli_group is not None and "." in cli_group:
+        raise ValueError("Blueprint CLI group names should not contain a dot.")
... 
+    if (
+        cli_resolved_group is not _sentinel
+        and cli_resolved_group is not None
+        and "." in cli_resolved_group
+    ):
+        raise ValueError("Blueprint CLI group names should not contain a dot.")
...
-    if endpoint:
-        assert "." not in endpoint, "Blueprint endpoints should not contain dots"
-    if view_func and hasattr(view_func, "__name__"):
-       assert (
-            "." not in view_func.__name__
-       ), "Blueprint view function name should not contain dots"
+    if endpoint is not None:
+        if "." in endpoint:
+            raise ValueError("Blueprint endpoints should not contain a dot.")
+    elif view_func and hasattr(view_func, "__name__") and "." in view_func.__name__:
+        raise ValueError(
+            "Blueprint view function name should not contain a dot"
+            " when an endpoint is not provided."
+        )
     self.record(lambda s: s.add_url_rule(rule, endpoint, view_func, **options))
\end{lstlisting}

% The patch refiner improves upon the initial draft in several critical ways:

% \begin{enumerate}
%     \item \textbf{Validates CLI Group Names}: The refiner adds checks for the \texttt{cli\_group} parameter in two places: during the Blueprint's initialization and within the \texttt{register} method. This prevents the creation of CLI command groups with dots in their names, an edge case completely missed by the first patch.
    
%     \item \textbf{Improves Endpoint and View Function Validation}: The original code used \texttt{assert} statements to check for dots in endpoint and view function names. The refined patch replaces these with explicit \texttt{ValueError} exceptions, which is better practice for validating user input. It also introduces more nuanced logic: it first checks the \texttt{endpoint} name, and only if an endpoint is not explicitly provided does it check the \texttt{view\_func}'s name. This provides a more precise and helpful error message to the developer.
% \end{enumerate}

By enforcing dot restrictions consistently across all naming conventions, the refined patch offers a reliable, complete resolution of the issue—demonstrating the power of context-aware patch refinement.